\documentclass[prd,twocolumn,pdflatex,showpacs,superscriptaddress]{revtex4-1}
\usepackage{graphicx}
\usepackage{dcolumn}
\usepackage{bm}
\usepackage{gensymb}
\usepackage{color}
\usepackage{subfig}

\usepackage{amssymb}
\usepackage{amsmath}
\usepackage{mathrsfs}

\DeclareMathAccent{\frmarr}{\mathord}{letters}{"7E}

\usepackage{mathtools}
\begin{document}

\preprint{APS/123-QED}

\title{Properties of Finite Amplitude Electromagnetic Waves propagating in the Quantum Vacuum}

\author{Hedvika Kadlecov\'{a}}
\email{Hedvika.Kadlecova@eli-beams.eu}
\affiliation{Institute of Physics of the ASCR, ELI--Beamlines project, Na Slovance 2, 18221, Prague, Czech Republic}
\author{Sergei V. Bulanov}
\affiliation{Institute of Physics of the ASCR, ELI--Beamlines project, Na Slovance 2, 18221, Prague, Czech Republic}
\affiliation{National Institutes for Quantum and Radiological Science and Technology (QST), Kansai Photon Science Institute, 8--1--7 Umemidai, Kizugawa, Kyoto 619--0215, Japan}
\affiliation{A. M. Prokhorov Institute of General Physics of RAS, Vavilov Str.~38, Moscow 119991, Russia}
\author{Georg Korn}
\affiliation{Institute of Physics of the ASCR, ELI--Beamlines project, Na Slovance 2, 18221, Prague, Czech Republic}

\date{\today}

\begin{abstract} 
We study two counter-propagating electromagnetic waves in the vacuum within the framework of the Heisenberg-Euler formalism in quantum electrodynamics. We show that the non--linear field equations decouple for ordinary wave case and can be solved exactly. We solve the non--linear field equations assuming the solution in a form of a Riemann wave.  
We discuss the properties of the  nonlinear electromagnetic wave propagating in the quantum vacuum, such as the wave steepening, subsequent generation of high order harmonics and electromagnetic shock wave formation with electron--positron pair generation at the shock wave front.
\end{abstract}

\maketitle


The increasing availability of high power lasers raises interest in experimental observation and motivates theoretical studies of non--linear QED in the laser-laser scattering \cite{Mourou,Marklund,DiPizzaReview,BulanovEli, Tommasini,Parades,King,KogaBulanov,KarbsteinShaisultanov, GiesKarbsteinKohlfurstSeegert,Cherenkov}, scattering of the XFEL emitted photons \cite{Inada}, and the interaction of relatively long wavelength radiation with the X-ray photons \cite{Schlenvoigt,Shangai100PW,Heinzlliesfeld}, nonlinear laser--plasma interaction \cite{Shukla,Piazza1} and to complex problems on the boundary of non--linear QED, super intense lasers and plasma, nuclear and particle physics such as 
theoretical studying of MeV X-rays in a plasma that is generated by femtosecond laser pulses, the study of $\gamma-$induced nuclear reactions in plasma radiated by super-intense laser, or neutron production in laser plasma \cite{Karsch,Pretzler,Izumi}, non--linear cooperative electron--gamma--nuclear, multiphoton and other processes \cite{Glushkov,Glushkov1,Khetselius}, fotonuclear physics, nuclear fission and fusion with laser--heated clusters and plasmas \cite{Ledingham,Ditmire, Umstadter, Ledingham1,Lezhnin}. 

The results expected to be obtained will allow us to test extensions of the Standard Model, in which new particles can participate in the loop diagrams and provide a window into a new physics (see search for the process in X--ray region \cite{Inada}).

Such non--linear process of light-by-light scattering breaks the linearity of the Maxwell equations and is one of the oldest predictions of quantum electrodynamics QED. The photon--photon scattering in a vacuum occurs via the generation of virtual electron--positron pairs creation resulting in vacuum polarization \cite{Electrodynamics}. To investigate such process it is convenient to use Heisenberg--Euler approach in QED \cite{HeisenbergEuler,Electrodynamics,Gies}. This problem was addressed as birefringence effect when the speed of wave propagation depends on the wave polarization, \cite{Bialynicka,DittrichGies}. 
Birefringence was motivated by analogy with the effect in crystalography and means that the incoming light splits into two waves in the vacuum which serves as a medium: the ordinary wave and the extraordinary wave. The ordinary wave propagates parallel to the optic axis with polarization perpendicular to the optic axis and refractive index $n_{or}$. Extraordinary wave has polarization in the direction of the optic axis and has refractive index $n_{ex}$. For example, when unpolarized light enters an uniaxial birefringent material, it is split into two beams travelling different directions. The ordinary ray doesn't change direction while the extraordinary ray is refracted as it travels through the material. The magnitude of birefringence is given by $\Delta n=n_{or}-n_{ex}$. Birefringence was also studied in astrophysics, \cite{HattoriI, HattoriII}.

The vacuum polarization process results in decreasing the velocity of counter--propagating electromagnetic waves. 


High--order harmonics generation in vacuum was studied in \cite{PiazzaHarmonics} where the generation takes place in the strong field of two counterpropagating laser waves. The production of odd harmonics in vacuum was investigated in \cite{FedotovNarozhny, FedotovNarozhnyThird, Rosanov93}.


High-order harmonics produced by laser--solid target interaction for lower laser intensites have been widely studied in the past decades thanks to their potential applications in ultraviolet or x-ray source generation, in attosecond dynamics studying and even in plasma surface detection, \cite{BulanovPegoraro}, and \cite{Mendonca1,HuHarmonics}. Recently, the dynamics of two ultra-relativistic intense counter-propagating lasers interacting with a thin foil target is studied by using QED module included PIC simulations. Harmonics up to 5th order have been demonstrated. It shows that such harmonics are generated due to the QED effects, \cite{YuHarmonics}. The rates of harmonics radiation at the electron--positron annihilation in the field of a strong pump wave were recently investigated in \cite{Avetissian}.


In the Heisenberg--Euler approximation of QED, \cite{HeisenbergEuler}, the electromagnetic fields propagate in the dispersionless media whose refraction index depends on the electromagnetic field. This leads to the nonlinear response and the electromagnetic wave evolves into a configuration with gradient singularities \cite{FluidMechanics} leading to formation of a shock wave. The occurance of singularities in the Heisenberg--Euler theory is noticed in \cite{LutzkyToll} where a  particular solution of field equations from Heisenberg--Euler Lagrangian is obtained. In \cite{BohlRuhl}, the wave steepening is demonstrated by numerical integration of nonlinear QED equations in vacuum. The EM shock waves are known, see for example \cite{GaponovOstrovskii} and \cite{JeffreyKorobeinikov}.

Let us mention that collisions of two gravitational waves/electromagnetic waves is addressed in General Relativity (GR). It is a difficult problem to find the spacetime structure occuring after the collision of gravitational and/or electromagnetic waves due to the nonlinearity of the field equations in GR. The problem is simplified by specializing to impulsive and/or shock waves which are plane and homogeneous. Then exact solutions can be found. The Khan-Penrose \cite{Khan} and Bell-Szekeres \cite{BellSzekeres} solutions are ones of the most famous. Exact solution taking into account the presence of cosmological constant was found in \cite{Barrabes}.

Recently, we have presented and analyzed an analytical solution of the non--linear field equations in QED, in the Heisenberg-Euler approximation, describing the finite amplitude electromagnetic wave counter--propagating to the crossed electromagnetic field, \cite{KadlecovaKornBulanov2019}. 


In the present paper, we  widen our analysis by implementing a new method of solving the system of non--linear equations. We will show that the non--linear field equations decouple for the ordinary wave case when we look for the solution in the form of a simple wave \cite{Kadomtsev, KadomtsevKarpman,FluidMechanics}, and we solve the decoupled equations exactly. 
The resulting non--linear wave equation is solved by integration along the characteristics of the equation. We demonstrate in more detail that the solution describes high order harmonic generation, wave steepening and formation of a shock wave.

The paper is organized as follows. In Section \ref{sec:Lagrangian}, we review the Heisenberg--Euler formalism which will be useful in the paper. We  use the weak field approximation to the sixth order in the field amplitude to include four and six photon interactions.

In Section \ref{sec:DerivationEquations}, we derive the non--linear field equations, we add weak linear amplitude corrections and linearize the coefficients. 

Next, in Section \ref{sec:SolvingEquations}, we solve the field equations. We present new analytical method of solving such system of equations, we assume the solution in a form of a simple wave, we show that the system of equations decouple for the ordinary wave case and it can be solved exactly. The solution has a form of a non--linear wave without dispersion in the linear approximation.

In Section \ref{sec:AnalyzingEquations} we concentrate on analyzing the solution. We
analyze the solution by the method of characteristics and by perturbation method. We discuss properties of the solution, such as wave breaking, in detail, in the cae of self--similar solutions with uniform (homogenious) deformation.
  
The main results of the paper are summarized in concluding Section \ref{sec:conclusion}.

\section{\label{sec:Lagrangian} Heisenberg--Euler Lagrangian}
The Heisenberg--Euler Lagrangian in the weak field approximation is given by
\begin{equation}
\mathcal{L}_{HE}=\mathcal{L}_{0}+\mathcal{L}',\label{eq:Lagrangian}
\end{equation}
where $\mathcal{L}_{0}=-(1/16\pi)F_{\mu\nu}F^{\mu\nu}$ is the classical electromagnetic Lagrangian, $F_{\mu\nu}$ is the electromagnetic field tensor $F_{\mu\nu}=\partial_{\mu}A_{\nu}-\partial_{\nu}A_{\mu}$, $\mu,\nu={0,1,2,3}$; $A_{\mu}$ is the 4-vector of the electromagnetic field and $\mathcal{L}'$ is the radiation correction in the Heisenberg--Euler theory, \cite{HeylHernquist,Electrodynamics}. In the weak field approximation, the Heisenberg--Euler Lagrangian has a form
\begin{align}
\mathcal{L}'=\kappa_{1}\left\{4 \mathcal{F}^2 + 7 \mathcal{G}^2 + \frac{90}{315}\mathcal{F}\left[16 \mathcal{F}^2 + 13 \mathcal{G}^2\right]\right\},\label{eq:Lagrangiandash}
\end{align}
where $\kappa_{1}=e^4/360 \pi^2 {m}^4$, $\mathcal{F}$ and $\mathcal{G}$ are the Poincar\'{e} invariants, which are defined in terms of the field tensor $F_{\mu\nu}$,
\begin{align}
&\mathcal{F}=\frac{1}{4}F_{\mu\nu}F^{\mu\nu}=\tfrac{1}{2}({\bf B}^2-{\bf E}^2),\\
&\mathcal{G}=\frac{1}{4}F_{\mu\nu}{\overset{\star}{F}}{}^{\mu\nu}={\bf E}\cdot {\bf B},\\
&{\overset{\star}{F}}{}^{\mu\nu}=\frac{1}{2}\varepsilon^{\mu\nu\rho\sigma}F_{\rho\sigma},
\end{align}
where ${\bf E}$ and ${\bf B}$ are electric and magnetic fields, $\varepsilon^{\mu\nu\rho\sigma}$ being the Levi-Civita symbol in four dimensions and we use the units $c=\hbar=1$.

The Lagrangian $\mathcal{L}'$ can be used if $\omega \ll m$ and $E \ll E_{S}$, where $\omega$ is characteristic frequency of the radiation, the  field
\begin{equation}
E_{S}=m^2_{e}/e\; (m^2_{e}c^3/e\hbar),
\end{equation}
is the critical field (or Schwinger field) in QED, $m_{e}$ is the electron rest mass, $e$ is the electron elementary charge.

Expanding the Lagrangian (\ref{eq:Lagrangiandash}) in the series, we keep the terms to the third order in the field amplitude within the weak field approximation to describe the singular solutions. The contributions of the fourth order cancel each other in calculation of dispersive properties of the QED vaccum. The remaining contribution is of the same order as from the Heisenberg--Euler Lagrangian expansion to the sixth order in the fields. The first two terms on the right hand side in the Lagrangian (\ref{eq:Lagrangiandash}) describe four interacting photons and the last two terms correspond to six photon interaction.

The field invariants $\mathcal{F}=\mathcal{G}=0$ in the limit of co--propagating waves. 

The field equations are given by 
\begin{equation}
\partial_{\mu}(\partial\mathcal{L}'/\partial(\partial_{\mu}{\Phi}))-\partial{\mathcal{L}'}/\partial\Phi=0,
\end{equation}
where
\begin{equation}
\Phi=(-\phi,\bf{A}).
\end{equation}

The first pair of Maxwell field equations reads
\begin{align}
\nabla \cdot {\bf B}&=0,\nonumber\\
\nabla \times {\bf E}&=-{\partial_{t} {\bf B}}. \label{FirstMax}
\end{align} 

The second pair can be found by varying the Heisenberg--Euler Lagrangian (\ref{eq:Lagrangian}) which gives the field equations. The second pair of equations can be written as
\begin{align}
\nabla \times {\bf H}&=\partial_{t} {\bf D},\nonumber\\
\nabla \cdot {\bf D}&=0, \label{SecondMax}
\end{align} 
where
\begin{align}
{\bf D}&={\bf E}+4\pi{\bf P},\nonumber\\
{\bf H}&={\bf B}-4\pi{\bf M},\label{eq:DH}\\
{\bf P}&=\partial_{\bf E}{\mathcal{L}_{\textrm{HE}}},\nonumber\\
{\bf M}&=\partial_{{\bf B}}{\mathcal{L}_{\textrm{HE}}}\nonumber, 
\end{align}
and ${\bf P}$ and ${\bf M}$ are the electric and magnetic polarization vectors. The derivatives  are defined as
\begin{align}
\partial_{\bf E}&=(\partial_{E_{x}},\partial_{E_{y}},\dots),\nonumber\\
\partial_{\bf B}&=(\partial_{B_{x}},\partial_{B_{y}},\dots)\label{eq:EB}.
\end{align}

\section{\label{sec:DerivationEquations} Heisenberg--Euler field equations}
We work in the orthogonal coordinate system, $(x,y,z)$, where the two waves propagate along the $x-$axes.
For the ordinary wave case, we assume ${\bf E}=(0,0,E_{z})$ and ${\bf B}=(0,B_{y},0)$,
the simple case of non--vanishing components $E_{z}$ and $B_{y}$ in order to investigate the crossed field case, (${\bf E}\cdot{\bf B}=0$). Then the first equation comes from the set of equations (\ref{FirstMax}), the second equation was found by varying the Lagrangian (\ref{eq:Lagrangian}) according to the potential ${\bf A}$:
\begin{equation}
\partial_{t}B_{y}-\partial_{x}E_{z}=0,\label{eq:nonlinear1}
\end{equation}
\begin{align}
-\left[1+8\kappa_{1} E^2_{z}+4(E^2_{z}-B^2_{y})(\kappa_{1}-3\kappa_{2}E^2_{z})\right.&\nonumber\\
-6\left.\kappa_2(E^2_{z}-B^2_{y})^2\right]&\partial_{t}E_{z}\nonumber\\
+\left[1-8\kappa_{1} B^2_{y}+4(E^2_{z}-B^2_{y})(\kappa_{1}+3\kappa_{2}B^2_{y})\right.&\nonumber\\
-6\left.\kappa_2(E^2_{z}-B^2_{y})^2\right]&\partial_{x}B_{y}\nonumber\\
+4\left[2\kappa_{1}-3\kappa_{2}(E^2_{z}-B^2_{y})\right] E_{z}B_{y}(\partial_{t}B_{y}+\partial_{x}E_{z})=&0,\label{eq:nonlinearr}
\end{align}
where we denote $\kappa_{2}=180/315\kappa_{1}$, $E_{z}\equiv E$ and $B_{y}\equiv B$, and add weak linear amplitude corrections to the fields,
\begin{align}
E&=E_{0}+a(x,t),\nonumber\\
B&=B_{0}+b(x,t).\label{eq:EB}
\end{align}
The fields $E_{0}, B_{0}$ represent the constant electromagnetic background field, $a(x,t)$ and $b(x,t)$ are functions of $x$ and $t$. Using expressions (\ref{eq:EB}),  equations (\ref{eq:nonlinearr}) can be rewritten in a form
\begin{align}
\partial_{t}b(x,t)&=\partial_{x}a(x,t), \label{eq:abequations}\\
\alpha\,\partial_{t}a(x,t)&-\beta\,[\partial_{x}a(x,t)+\partial_{t}b(x,t)]-\gamma\,\partial_{x}b(x,t)=0,\label{eq:shift}
\end{align}
where the coefficients $\alpha, \beta$ and $\gamma$ are:
\begin{align}
\alpha&=1+8\kappa_{1} (E_{0}+a)^2\nonumber\\
+&4\left[(E_{0}+a)^2-(B_{0}+b)^2\right](\kappa_{1}-3(E_{0}+a)^2\kappa_{2})\\
-&6\kappa_{2}\left[(E_{0}+a)^2-(B_{0}+b)^2\right]^2,\nonumber\\
\beta&=4(E_{0}+a)(B_{0}+b)\left[2\kappa_{1}-3\kappa_{2}[(E_{0}+a)^2-(B_{0}+b)^2]\right],\\
\gamma&=1-8\kappa_{1} (B_{0}+b)^2\nonumber\\
+&4\left[(E_{0}+a)^2-(B_{0}+b)^2\right](\kappa_{1}+3(B_{0}+b)^2\kappa_{2})\label{eq:ABC}\\
-&6\kappa_{2}\left[(E_{0}+a)^2-(B_{0}+b)^2\right]^2.\nonumber
\end{align}
Assuming that $a(x,t)=b(x,t)=0$, and the crossed field case $E_{0}=B_{0}$, we obtain that 
\begin{align}
\alpha_{0}&=1+8\kappa_{1} E^2_{0},\nonumber\\
\beta_{0}&=8\kappa_{1} E^2_{0},\label{eq:ABCcrossed}\\
\gamma_{0}&=1-8\kappa_{1} E^2_{0}.\nonumber
\end{align}

To find the wave phase velocity from the linearized equations (\ref{eq:abequations}) and (\ref{eq:shift}) we look for the solutions in the form,
\begin{equation}
a \propto \exp(-i\omega t + i qx),\;\; b \propto \exp(-i\omega t + i qx), \label{eq:ab}
\end{equation}
where $q$ is the wave number and $\omega$ is the frequency.
Substituting (\ref{eq:ab}) into the equations (\ref{eq:abequations}) and (\ref{eq:shift}), and dividing them by wave vector $q$, we obtain algebraic set of equations for the wave velocity $v={\omega}/q$ (since the medium is dispersionless in our study, see Eq.~(\ref{eq:finalResult})), we denote the phase and the group velocity as one $v=v_{ph}=v_{g}; v_{ph}=\omega/q, v_{g}=\partial{\omega}/\partial{q}$. It yields equations
\begin{align}
a+vb&=0,\nonumber\\
v(b\beta_{0}-a\alpha_{0})-(a\beta_{0}+b\gamma_{0})&=0, \label{eq:vphasetwo}
\end{align}
whose solution is
\begin{align}
v_{1,2}&=\frac{-\beta_{0}\pm \sqrt{\beta^2_{0}+\alpha_{0}\gamma_{0}}}{\alpha_{0}}.\label{eq:vphsolutionExtra}
\end{align}
Using  relationships given by Eq.~(\ref{eq:ABCcrossed}) we find 
\begin{align}
v_{1}&=-1,\nonumber\\
v_{2}&=\frac{\gamma_{0}}{\alpha_{0}}=\frac{1-8\kappa_{1} E^2_{0}}{1+8\kappa_{1} E^2_{0}}. \label{eq:vphasetwo}
\end{align}
This is the phase velocity $v=v_{1,2}$ for the wave propagating over the crossed background field in the weak field approximation of the Heisenberg--Euler theory. Similar problem is studied in \cite{Bialynicka, Marklund, DittrichGies} and \cite{Rozanov} where the strong static homogeneous background field is considered. The obtained result is used further as a limit case for the background crossed field.

We assuming the coefficients $\alpha, \beta$ and $\gamma$ in the form:
\begin{align}
\alpha&=\alpha_{0}+\alpha_{a}a+\alpha_{b}b,\nonumber\\
\beta&=\beta_{0}+\beta_{a}a+\beta_{b}b,\label{eq:ABCdelta}\\
\gamma&=\gamma_{0}+\gamma_{a}a+\gamma_{b}b,\nonumber
\end{align}
where 
\begin{align}
\alpha_{a}&=(\partial_{a} {\alpha})|_{a=0},\quad
\alpha_{b}=(\partial_{b}{\alpha})|_{b=0},\nonumber\\
\beta_{a}&=(\partial_{a}{\beta})|_{a=0},\quad
\beta_{b}=(\partial_{b}{\beta})|_{b=0},\label{eq:alphabeta}\\
\gamma_{a}&=(\partial_{a}{\gamma})|_{a=0},\quad
\gamma_{b}=(\partial_{b}{\gamma})|_{b=0}.\nonumber
\end{align}

We can identify the coefficients $\alpha_{a}, \beta_{a}, \gamma_{a}$ and $\alpha_{b}, \beta_{b}, \beta_{b}$, with the general form of $\alpha, \beta, \gamma$ (\ref{eq:ABC}) for the crossed field $E_{0}=B_{0}$. It yields
\begin{align}
\alpha_{a}&=24 E_{0}(\kappa_{1} - \kappa_{2} E^2_{0})+48\kappa_{2}E_{0}(b^2+2E_{0}b)|_{a=0},\nonumber\\
\alpha_{b}&=-8E_{0}(\kappa_{1} - 3\kappa_{2}E^2_{0})+48\kappa_{2}E_{0}(a^2+2E_{0}a)|_{b=0},\nonumber\\
\beta_{a}&=4(E_{0}+b|_{a=0}) \left[2\kappa_{1}-6\kappa_{2}E^2_{0}-3\kappa_{2}(b^2+2E_{0}b)|_{a=0}\right],\nonumber\\
\beta_{b}&=4(E_{0}+a|_{b=0}) \left[2\kappa_{1}+6\kappa_{2}E^2_{0}-3\kappa_{2}(a^2+2E_{0}a)|_{b=0}\right],\nonumber\\
\gamma_{a}&=8E_{0}(\kappa_{1}+3\kappa_{2}E^2_{0})+48\kappa_{2}E_{0}(b^2+2E_{0}b)|_{a=0},\label{eq:koef}\\
\gamma_{b}&=-24E_{0}(\kappa_{1}+\kappa_{2}E^2_{0})+48\kappa_{2}E_{0}(a^2+2E_{0}a)|_{b=0}.\nonumber
\end{align}

The terms 
\begin{align}
b|_{a=0}&=0,\quad
a|_{b=0}=0, \label{ordinary:conditions}
\end{align}
should be equal to zero because they are not linear and break the linear approximation we assume. We will use these conditions (\ref{ordinary:conditions}) to specify the constant in Eq.~(\ref{eq:finalOrdinary2}) while solving the non--linear equations.

\section{\label{sec:SolvingEquations} Self--similar solutions}

First, we consider the equations (\ref{eq:abequations}, \ref{eq:shift}) for the ordinary wave with functions $\alpha(a,b), \beta(a,b)$ and $\gamma(a,b)$ (\ref{eq:ABC}) in linear approximation (\ref{eq:ABCdelta}). We solve the non--linear equations using simple wave concept (Riemann wave) known in nonlinear wave theory \cite{KadomtsevKarpman, Kadomtsev, Whitham}. 

Equivalently, we assume the dependence $b=b(a)$,
and subsequently $\partial_{t}b=({\rm d} b/{\rm d} a)\partial_{t}a,\;\partial_{x}b=({\rm d} b/{\rm d} a)\partial_{x}a$. Eqs.~(\ref{eq:abequations}, \ref{eq:shift}) become
\begin{align}
\partial_{t}a&=\frac{{\rm d} a}{{\rm d} b} \partial_{x}a,\label{eq:partOne}\\
\partial_{t}a&=\frac{1}{\alpha}\left(2\beta+\gamma\frac{{\rm d} b}{{\rm d} a}\right) \partial_{x}a,\label{eq:partTwo}
\end{align}
while comparing the two equations, we obtain a quadratic equation for function $b(a)$. It has a form
\begin{align}
\gamma\left(\frac{{\rm d}b}{{\rm d}a}\right)^2+2\beta\frac{{\rm d}b}{{\rm d}a}-\alpha=0,\label{eq:quadraticOrdinary}
\end{align}
and has two solutions
\begin{align}
\left(\frac{{\rm d}b}{{\rm d}a}\right)=\frac{-\beta\pm \sqrt{\beta^2+\alpha\gamma}}{\gamma}.\label{eq:solutionsI}
\end{align}

We use a weak but finite amplitude approximation, assuming that the solution has a form
\begin{align}
\left(\frac{{\rm d}b}{{\rm d}a}\right)=\nu,\quad \nu=\nu_{0}+\nu_{a}a+\nu_{b}b.\label{eq:linearSolutionsOO}
\end{align}

For the calculation, we use the definition of tangent to a surface at a point $(\alpha_{0}, \beta_{0},\gamma_{0})$ as 
\begin{align}
f(\alpha,\beta,\gamma)&=f(\alpha,\beta,\gamma)|_{\alpha_{0},\beta_{0},\gamma_{0}}+\partial_{\alpha}{f}|_{\alpha_{0},\beta_{0},\gamma_{0}}(\alpha-\alpha_{0})\nonumber\\
+&\partial_{\beta}{f}|_{\alpha_{0},\beta_{0},\gamma_{0}}(\beta-\beta_{0})+\partial_{\gamma}{f}|_{\alpha_{0},\beta_{0},\gamma_{0}}(\gamma-\gamma_{0}),
\end{align}
where ${{\rm d}b}/{{\rm d}a}=f(\alpha,\beta,\gamma)$.
As a results we obtain coefficients
\begin{align}
\nu_{0}=&f|_{\alpha_{0},\beta_{0},\gamma_{0}}=\frac{-\beta_{0}\pm 1}{\gamma_{0}},\label{eq:nu0}\\
\partial_{\alpha}{f}|_{\alpha_{0},\beta_{0},\gamma_{0}}&=\pm\frac{1}{2},\label{eq:nu1}\\
\partial_{\beta}{f}|_{\alpha_{0},\beta_{0},\gamma_{0}}&=\frac{1}{\gamma_{0}}\left(-1\pm\beta_{0}\right),\label{eq:nu2}\\
\partial_{\gamma}{f}|_{\alpha_{0},\beta_{0},\gamma_{0}}&=\pm\frac{\alpha_{0}}{2\gamma_{0}}-\frac{\left(-\beta_{0}\pm 1 \right)}{\gamma_{0}^2},\label{eq:nu3}
\end{align}
and $\alpha-\alpha_{0}=\alpha_{a}a+\alpha_{b}b$, $\beta-\beta_{0}=\beta_{a}a+\beta_{b}b$ and $\gamma-\gamma_{0}=\gamma_{a}a+\gamma_{b}b$  where we have used $\beta^2_{0}+\alpha_{0}\gamma_{0}=1$.

The complete set of linear coefficients in (\ref{eq:linearSolutionsOO}) is
\begin{align}
\nu_{0}=&f|_{\alpha_{0},\beta_{0},\gamma_{0}},\nonumber\\
\nu_{a}=&\alpha_{a}f_{\alpha}+\beta_{a}f_{\beta}+\gamma_{a}f_{\gamma},\label{eq:nu}\\
\nu_{b}=&\alpha_{b}f_{\alpha}+\beta_{b}f_{\beta}+\gamma_{b}f_{\gamma},\nonumber
\end{align}
where the derivatives are denoted
\begin{align}
f_{\alpha}=&\partial_{\alpha}{f}|_{\alpha_{0},\beta_{0},\gamma_{0}},\;
f_{\beta}=\partial_{\beta}{f}|_{\alpha_{0},\beta_{0},\gamma_{0}},\;
f_{\gamma}=\partial_{\gamma}{f}|_{\alpha_{0},\beta_{0},\gamma_{0}}.\label{eq:fff}
\end{align}

Since we have two solutions of the equation (\ref{eq:solutionsI}) we need to choose the physical one, which corresponds to the case of two counter propagating waves. We can do that by knowing the phase velocity for such case, phase velocity $v=v_{2}>0$ (\ref{eq:vphasetwo}), and expression for $\nu_{0}$ (\ref{eq:nu0}). It shows that we need to choose the $-$ solutions, the $+$ solutions correspond to two waves propagating in the same direction.

Therefore evaluating $f_{\alpha}, f_{\beta}, f_{\gamma}$ (\ref{eq:fff}) by using expressions (\ref{eq:nu0}), (\ref{eq:nu1}), (\ref{eq:nu2}) and (\ref{eq:nu3}), we get
\begin{align}
f_{\alpha}&=-\frac{1}{2},\nonumber\\
f_{\beta}&=-\frac{1}{\gamma_{0}}\left(1 + \beta_{0}\right),\label{eq:ff}\\
f_{\gamma}&=-\frac{\alpha_{0}}{2\gamma_{0}}+\left(\frac{\beta_{0}+1}{\gamma^2_{0}}\right).\nonumber
\end{align}

Now, we observe that the problem reduces to solving the differential equation (\ref{eq:linearSolutionsOO}). The equation is in a form of total differential. It can be solved by the method of integration factor, choosing it as $m(a)=\exp(-\nu_{b}a)$. The dependence $b=b(a)$ is
\begin{equation}
\frac{1}{\nu_{b}}\exp{(-\nu_{b}a)}\left((\nu_{0}+\nu_{b}b)+\frac{\nu_{a}}{\nu_{b}}(\nu_{b}a+1)\right)=\delta,\label{eq:solutionImplicit}
\end{equation}
where $\delta$ is arbitrary constant. Therefore the function $b=b(a)$ has a form
\begin{equation}
b=\delta\,\exp(\nu_{b}a)-\frac{\nu_{a}}{\nu_{b}}(\nu_{b}a+1)-\frac{\nu_{0}}{\nu_{b}}.\label{eq:solutionExplicit}
\end{equation}

The remaining constant $\delta$ can be determined by the conditions (\ref{ordinary:conditions}) and therefore it allows one to find the constant,
\begin{equation}
\delta=\frac{\nu_{a}+\nu_{0}\nu_{b}}{\nu^2_{b}}.\label{eq:constOrdinary}
\end{equation}

Then the coefficients (\ref{eq:koef}) get a final form
\begin{align}
\alpha_{a}&=24 E_{0}(\kappa_{1} - \kappa_{2} E^2_{0}),\;
\alpha_{b}=-8E_{0}(\kappa_{1} - 3\kappa_{2}E^2_{0}),\nonumber\\
\beta_{a}&=8E_{0} \left[\kappa_{1}-3\kappa_{2}E^2_{0}\right],\;
\beta_{b}=8E_{0} \left[\kappa_{1}+3\kappa_{2}E^2_{0}\right],\label{eq:koeffinal}\\
\gamma_{a}&=8E_{0}(\kappa_{1}+3\kappa_{2}E^2_{0}),\; \gamma_{b}=-24E_{0}(\kappa_{1}+\kappa_{2}E^2_{0}).\nonumber
\end{align}
In order to use the weak amplitude approximation, we perform Taylor expansion of the first term in (\ref{eq:solutionExplicit}) to the first order, $\exp{(\nu_{b}a)}\approx 1+\nu_{b}a+\dots$ and it gives
\begin{equation}
b=\delta\,(\nu_{b}a+1)-\frac{\nu_{a}}{\nu_{b}}(\nu_{b}a+1)-\frac{\nu_{0}}{\nu_{b}}.\label{eq:solutionExplicitExp}
\end{equation}
After substituting (\ref{eq:constOrdinary}) into (\ref{eq:solutionExplicitExp}), we obtain the solution showing a linear relationship between $a$ and $b$:
\begin{equation}
b=\nu_{0}a,\label{eq:baOrdinary}
\end{equation}
where
\begin{equation}
\nu_{0}=-1/v,\label{eq:nuphase}
\end{equation}
and $v$ is the phase velocity  (\ref{eq:vphsolutionExtra}).

Now, we will get back to the equations (\ref{eq:partOne}) and (\ref{eq:partTwo}). 
It is more convenient to use Eq.~(\ref{eq:partOne}), which we rewrite as
\begin{equation}\label{eq:finalOrdinary}
\partial_{t}a-\frac{1}{\nu}\partial_{x}a=0,
\end{equation}
where it is denoted 
\begin{equation}
\nu=\nu_{0}+\nu_{a}a+\nu_{b}b.
\end{equation}

We perform another linearization of $1/\nu$ as 
\begin{align}
f(\nu)&=f(\nu)|_{\nu_{0}}+\partial_{\nu}{f}|_{\nu_{0}}(\nu-\nu_{0}),
\end{align}
and obtain
\begin{align}
\frac{1}{\nu}=\frac{1}{\nu_{0}}\left(1-a\frac{\nu_{a}+\nu_{0}\nu_{b}}{\nu_{0}}\right).\label{eq:1nu}
\end{align}

It is convenient to rewrite Eq.~(\ref{eq:finalOrdinary}) with $1/\nu$ (\ref{eq:1nu}) to a final form:
\begin{equation}\label{eq:finalOrdinary2}
\partial_{t}a+f(a)\partial_{x}a=0,
\end{equation}
with
\begin{equation}\label{eq:finalResult}
f(a)=-\frac{1}{\nu_{0}}\left[1-a\frac{(\nu_{a}+\nu_{0}\nu_{b})}{\nu_{0}}\right]
\end{equation}
or
\begin{equation}\label{eq:finalResult1}
f(a)=v+a\frac{(\nu_{a}+\nu_{0}\nu_{b})}{\nu_{0}^2},
\end{equation}
where $v$ is the phase velocity of the electromagnetic wave.

The equation (\ref{eq:finalOrdinary2}) can be rewritten for the function
\begin{equation}
\bar{a}=\frac{(\nu_{a}+\nu_{0}\nu_{b})}{\nu_{0}^2} a,\label{eq:koef}
\end{equation}  
in a standard form, \cite{KadomtsevKarpman,Kadomtsev}, 
\begin{equation}\label{eq:finalOrdinary22}
\partial_{t}\bar{a}+(v+\bar{a})\partial_{x}\bar{a}=0.
\end{equation}

This is the final equation, which we analyze further. The form of the equation (\ref{eq:finalOrdinary2}) corresponds to the equation of non--linear wave without dispersion 
\cite{Kadomtsev}. The wave steepening takes place. The ordinary wave overturns as we demonstrate in detail together with the higher--order harmonics analysis in the next Section \ref{sec:AnalyzingEquations}.
 In the limit $a=0$, the wave moves with the phase velocity for the unperturbed case. 


\section{\label{sec:AnalyzingEquations} Properties of self--similar solutions}
In this Section, the equation (\ref{eq:finalOrdinary2}) is analyzed.

\subsection{\label{sub:char} Analyzing the equations using characteristics}
The equation (\ref{eq:finalOrdinary2}) can be solved by method of characteristics. Characteristic equations for Eq.~(\ref{eq:finalOrdinary2}) are
\begin{equation}
\frac{{\rm d}x}{{\rm d}t}=f(a),\; \frac{{\rm d}a}{{\rm d}t}=0.
\end{equation} 
Their solutions are $a(x,t)=A_{0}(x_{0})$ and $x=f(A_{0}(x_{0}))t+x_{0}$. The function $a(x,t)$ transfers along the characteristic $x_{0}$ without any distortion. Therefore for any differentiable function $A=A(x)$ we can write solution $a$ in a form
\begin{equation}
a(x,t)=A_{0}(x_{0})=A_{0}[x-f(a(x,t))t],\label{eq:A}
\end{equation}
where $A_{0}$ is an arbitrary function determined by initial condition, $a(x)|_{t=0}=A_{0}(x)$. We will choose the arbitrary function as $a(x,t)=A_{0}(x_{0})=a_{m}\sin(kx_{0})$ giving
\begin{equation}
a(x,t)=a_{m}\sin{[k(x-f(a(x,t))t)]}.\label{eq:axt}
\end{equation}

\subsection{\label{sub:wave} The wave breaking}
The wave breaking is typical behavior of waves in nonlinear dispersionless media. The solution of equation (\ref{eq:finalOrdinary2}) can be written in a implicit form (\ref{eq:A}) with the Euler coordinate $x$ dependent on the Lagrange coordinate $x_{0}$ and time.
The location where the wave breaks is determined by the gradient of function $a(x,t)$, the wave breaks when gradient becomes infinite, \cite{Panchenko}. We obtain such result by deriving (\ref{eq:A}), as
\begin{align}
\partial_{x}a=\frac{A'_{0}(x_{0})}{1+A'_{0}(x_{0})f'\,t},\quad t_{br}=-\frac{1}{A'_{0}(x_{0})f'},\label{eq:gradient}
\end{align}
where it is denoted $A'(x_{0})=\rm{d}A_{0}/\rm{d}x_{0}$ and $f'=\partial_{a}f(a)$.
The gradient becomes infinite at time $t_{br}$ when the denominator of (\ref{eq:gradient}) vanishes at some point $x_{br}$. At the time $t_{br}$ when the wave breaks the velocity $a_(x_br,t_{br})$ remains constant. Such singularity is called the wave breaking or the gradient catastrophe.

By using our ansatz for the solution (\ref{eq:axt}) we obtain 
\begin{align}
A'_{0}(x_{0})&=a_{m}k\cos{(kx_{0})},\\
f'&=\frac{\nu_{a}+\nu_{0}\nu_{b}}{\nu^2_{0}},
\end{align}
therefore
the gradient (\ref{eq:gradient}) and the wave breaking time $t_{br}$ result in
\begin{align}
\partial_{x}a=\frac{a_{m}k\cos{(kx_{0})}}{1+a_{m}k f'\,\cos{(kx_{0})}t},\nonumber\\
t_{b_{br}}=-\frac{1}{a_{m}k\cos{(k x_{0})} f'}, \label{eq:time}
\end{align}
and at the coordinate $x_{0}$ where $a_{m}k\cos{kx_{0}}$ is maximal, the velocity $a(x_{br},t_{br})=a_{m}\sin{[k(x_{br}-f(a(x_{br},t_{br}))]}$ remains constant. In \cite{KadlecovaKornBulanov2019}, we have showed the wave steepening evolving in time. Here, we will concentrate on investigation of the direction of the wave breaking in detail. The direction of the wave breaking depends on the sign of $f'$ in (\ref{eq:time}) which we discuss in the next subsection.

\subsection{\label{sub:char} Analyzing character of wave breaking}

We need to investigate the expression for 
\begin{equation}
\bar{a}=f' a.\label{eq:koefi}
\end{equation}

The resulting electromagnetic wave propagates along the $x-$coordinate according to (\ref{eq:finalOrdinary2}) and the direction of the wave breaking is given by the sign in front of function $f_{1}$.

As noted above, $f_{0}=v>0$, which is the phase velocity of the background field. 

By susbtituting $\alpha_{0}$, $\beta_{0}$ and $\gamma_{0}$ (\ref{eq:ABCcrossed}) into $f_{\alpha}, f_{\beta}, f_{\gamma}$  (\ref{eq:ff}) we observe it is convenient to express the functions in terms of the phase velocity $v$. It yields
\begin{align}
f_{\alpha}&=- \frac{1}{2},\;
f_{\beta}=-\frac{1}{v},\label{eq:ff11}\;
f_{\gamma}=\frac{1}{2}\frac{1}{v^2}.\nonumber
\end{align}

Then the coefficients $\nu_{a}, \nu_{b}$ are
\begin{align}
\nu_{a}&=\frac{4E_{0}}{v^2}\left[\kappa_{1}(1-2v-3v^2)+3\kappa_{2}E^2_{0}(v^2+2v-1)\right],\nonumber\\
\nu_{b}&=-\frac{4E_{0}}{v^2}\left[\kappa_{1}(3-2v-v^2)+3\kappa_{2}E^2_{0}(v-1)^2\right].
\end{align}

The function $f_{1}$ (\ref{eq:ff}) becomes
\begin{align}
f'=4E_{0}&\left[\kappa_{1}\left(-1-3v-3v^2+\frac{3}{v}\right)\right.\nonumber\\
+&3\left.\kappa_{2}E^2_{0}\left(v^2+3v-3+\frac{1}{v}\right)\right],\label{eq:f1res}
\end{align}
where $v=v_{2}$ (\ref{eq:vphasetwo}) as
\begin{equation}
v=\frac{1-8\kappa_{1} E^2_{0}}{1+8\kappa_{1} E^2_{0}}.\label{eq:velocity}
\end{equation}
We have obtained more general formula for the steepening factor $f'$ than in \cite{KadlecovaKornBulanov2019}, where $f'=-2(4\epsilon^2_{2}+3\epsilon_{3})W^3$, $W^3=-2\sqrt{2}E^3_{0}$ and $\epsilon_{2}=8\kappa_{1}$. 
If we substitute a Taylor expansion of $(\ref{eq:velocity})$ as 
\begin{equation}
v\approx 1-16\kappa_{1}E^2_{0}+16\kappa^2_{1}E^4_{0}, \label{eq:Taylor}
\end{equation}
into (\ref{eq:f1res}) and look for the terms with $E^3_{0}$,
we obtain
\begin{equation}
f'=48E^3_{0}[12\kappa^2_{1}+\kappa_{2}],
\end{equation}
which corresponds to the result in \cite{KadlecovaKornBulanov2019} where the wave has rarefaction character.

We can rewrite the function $f'$ in a final form as
\begin{align}
f'=&\frac{4E_{0}}{v}\left\{3(\kappa_{1}+\kappa_{2}E^2_{0})-v(\kappa_{1}+9\kappa_{2}E^2_{0})\right.\nonumber\\
-&3v^2\left.(\kappa_{1}-9\kappa_{2}E^2_{0})-3v^3(\kappa_{1}-\kappa_{2}E^2_{0})\right\},\label{eq:f1explicit}
\end{align}
where the phase velocity $v < 1$ and  the constants $\kappa_{1}=\alpha/360\pi^2\times 1/E^2_{S}$ and $\kappa_{2}=\kappa_{1}\times 180/315$ and $\alpha=1/137$. The constants without the scaling factor $1/E^2_{S}$  have values $\kappa_{1}\approx 2\times 10^{-6}$ and $\kappa_{2}\approx 10^{-6}$.



When the singularity is formed, the electromagnetic wave breaking formes a shock wave, which has a forward character for $f'>0$ and rarefaction character, i.e. the wave breaks in the backwards direction, for $f'<0$.

\begin{figure}[h]
\centering
\includegraphics[width=0.515\textwidth]{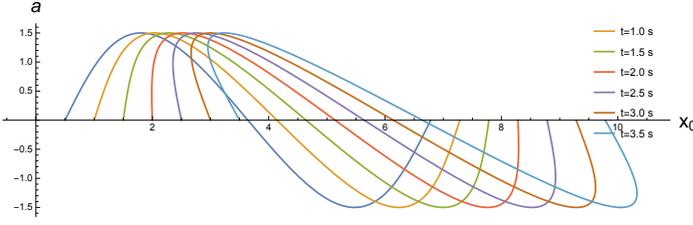}
\caption{\label{fig:shift} The equation (\ref{eq:image}) is visualized, the shifting of the wave in time to the left hand side is visible. }
\end{figure}
\begin{figure}[h]
\centering
\includegraphics[width=0.5\textwidth]{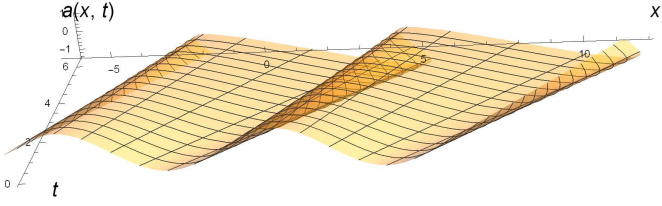}
\caption{\label{fig:shift3D} The equation (\ref{eq:image}) is visualized, the shifting of the wave in time to left hand side is visible in 3D. }
\end{figure}

The rarefaction character of the wave steepening is shown in Fig.~\ref{fig:shift}, where we plot 
\begin{equation}
x=x_{0}+(1+f'a_{0}(x_{0}))t,\quad a_{0}(x_{0})=a_{m}\sin(x_{0}),\label{eq:image}
\end{equation}
 for $f'=-0.35$ and $a_{m}=1$. The wave front shifts to left gradually. The situation in 3D is shown in Fig.~\ref{fig:shift3D} for $x_{0}\in<-2\pi, 2\pi>$.

\subsection{\label{sub:pert} Analyzing the equations by perturbation method}
Another way how to describe the wave breaking is to use the perturbation method \cite{Panchenko, Kadomtsev} to find the solutions of the equation (\ref{eq:finalOrdinary2}). We can write the solution as
\begin{equation}
a=a^{(0)}+\varepsilon a^{(1)}+\varepsilon^2 a^{(2)}+\dots,\label{per}
\end{equation} 
where $\varepsilon\ll 1$ and we assume that in the zeroth order the wave amplitude is homogeneous with the velocity, $a^{(0)}$, and constant in space and time.

In the first order $\varepsilon^{0}$, we obtain
\begin{equation}
\partial_{t}a^{(1)}+f(a)|_{a=0}a^{(1)}=0,\label{eq:first}
\end{equation} 
where $f(a)|_{a=0}=-1/\nu_{0}=v$, (\ref{eq:nuphase}).  We have obtained the simplest wave equation describing the wave with the frequency and wave number related via dispersion relation $\omega=k f(a)|_{a=0}$. Therefore we obtained that the wave propagates without dispersion and both the phase velocity, $\omega/k$, and the group velocity $\partial \omega/ \partial k$ are equal to $f(a)|_{a=0}=v$.

The solution of Eq.~(\ref{eq:first}) is arbitrary function of $x-vt$, where $v=f(a)|_{a=0}$, therefore we choose the same form as before,
\begin{equation}
a^{(1)}=a_{m}\sin{[k(x-vt)]}.
\end{equation}

To the second order, $\varepsilon^{1}$, we obtain
\begin{align}
\partial_{t}a^{(2)}+3v\partial_{x}a^{(2)}&=-a^{(1)}f'\partial_{x}a^{(1)},\nonumber\\
&=-\frac{bk}{2}\sin{[2k(x-vt)]},\label{eq:secondorder}
\end{align}
where $b=a^2_{m}f'$. The solution of this equation,
\begin{equation}
a^{(2)}=\frac{b}{8v}\left[\cos{[2k(x-vt)]}-\cos{(2kx)}\right],
\end{equation}
where we assumed $a^{(2)}|_{t=0}=0$, describes the second harmonic with the resonant growth of amplitude in time. In the third order, $\varepsilon^{(2)}$, we will find that the third harmonic grows as $\sin{[3k(x-vt)]}$ and so on. 
In general, the harmonics spectrum in the expansion can be estimated as
\begin{equation}
a_{n}=\left(f'\frac{E_{pulse}}{E_{S}}\right)^n,
\end{equation}
where $n$ is the order of the harmonic, $E_{pulse}$ is a typical field of the electromagnetic pulse which is not scaled with $E_{S}$.

We have demonstrated that the second harmonic is in resonance with the first harmonic and the two counter propagating electromagnetic waves propagate in vacuum without dispersion. The higher harmonics are generated up to the point of wave overturning.

\subsection{\label{sub:pert2} Analyzing the solution by perturbation method}
It is possible to directly analyze the solution (\ref{eq:axt}), which is in implicit form, by perturbation method. We assume the form of solutions as (\ref{per}) and we rewrite function $f(a)$ as
\begin{equation}
f(a)=f_{0}+\varepsilon a, \label{eq:favar}
\end{equation}
and 
\begin{equation}
a(x,t)=a_{m}\sin{[k(x-(f_{0}+\varepsilon a(x,t))t)]},\label{eq:axti}
\end{equation}
where $\varepsilon=f'$ and $f_{0}=-1/\nu_{0}$. In the first order $\varepsilon^{0}$, we obtain 
\begin{equation}
a_{0}=a_{m}\sin{[k(x-f_{0}t)]}.
\end{equation}
In the second order $\varepsilon^{1}$ and using Taylor expansion on right hand side, we
obtain 
\begin{equation}
a_{1}=-\frac{a^2_{m}kt}{2}\sin{(2kx_{0})},
\end{equation}
which describes the second harmonic with the resonant growth of amplitude in time. Again, we have demonstrated the resonance between the first two harmonics, which is true for all harmonics because the phase velocity is the same for all harmonics and it does not depend on the wave number. This again leads to wave breaking.\\

\subsection{\label{sub:deformationExample} Self--similar solutions with uniform deformation}
We assume the solution $a(x,t)$ of Eq.~(\ref{eq:finalOrdinary2}) in the form
\begin{equation}
a(x,t)=a_{0}(t)+a_{1}(t)x.\label{eq:triangular}
\end{equation}
It represents a triangular shape of the solution $a(x,t)$.
The function $f(a)$ (\ref{eq:finalResult}) can be rewritten as
\begin{equation}
f(a)=f_{0}+f' a, f_{0}=-\frac{1}{\nu_{0}}, f'=\frac{\nu_{a}+\nu_{0}\nu_{b}}{\nu^2_{0}}.
\end{equation}
After substituting the solution (\ref{eq:triangular}) into the equation (\ref{eq:finalOrdinary2}), we obtain the set of equations:
\begin{align}
\partial_{t}a_{0}+a_{1}(f_{0}+f'a_{0})&=0,\nonumber\\
\partial_{t}a_{1}+f'a^2_{1}&=0.\label{eq:Ex2}
\end{align}
The profile $a_{1}$ of the solution can be investigated by solving the second Eq.~(\ref{eq:Ex2}) as
\begin{equation}
a_{1}=\frac{a_{1}(0)}{1+f'a_{1}(0)t},\label{eq:steep}
\end{equation}
where $a_{1}(0)=a_{1}(t)|_{t=0}$.

\begin{figure}[!tbp]
\centering
\subfloat[The rarefaction wave breaks.]{\includegraphics[width=0.43\textwidth]{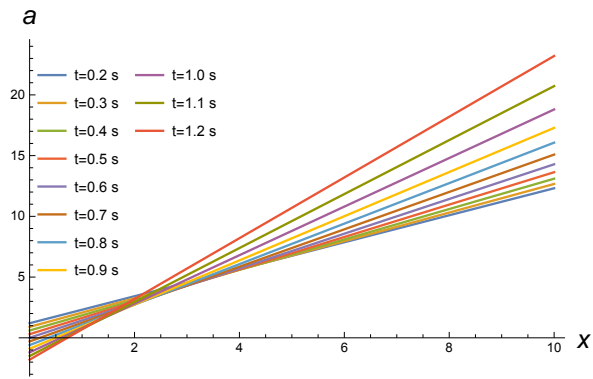}\label{fig:ex1}}
\hfill
\subfloat[The rarefaction waves do not break.]{\includegraphics[width=0.45\textwidth]{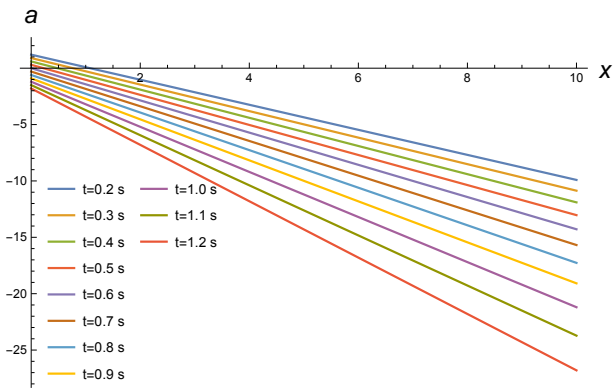}\label{fig:ex2}}
\caption{The equation (\ref{eq:triangular}) together with solution (\ref{eq:steep}) is visualized. In the Fig.~\ref{fig:ex1}, the shifting of the wave to left hand side is visible in time. We have chosen the function $a_{1}(0)=\cos(0)=1>0$, $a_{0}(t)=-3t+1.8$, $f'=-0.5$. In the Fig.~\ref{fig:ex2}, we have just changed the function $a_{1}(0)=\cos(\pi)=-1<0$ and $f'=0.5$ to positive value. The wave does not break and continues to infinity in time.}
\end{figure}

We can analyze the profile, for $a_{1}(0)>0$, and for $f'<0$, $a_{1} \rightarrow \infty$ and $t \rightarrow -1/f'a_{1}(0)$. The wave steepens to the left hand side in the opposite direction than the direction of propagation along the positive $x$ axes, i.e. has a rarefaction character, such behaviour is showed in Fig.~\ref{fig:ex1}. 

If $a_{1}(0)<0$, and for $f'<0$, $a_{1} \rightarrow 1/f't$ and $t \rightarrow \infty$. The wave does not break and continues till infinity, such behaviour is showed in Fig.~\ref{fig:ex2}.
For the case when $f'>0$, the direction of propagation just changes to the opposite direction.

\section{\label{sec:discussion} Dissipation due to the electron--positron pair creation}
Our work is performed within the approximation of Heisenberg--Euler theory of QED in the low photon energy region $\omega\ll m$, i.e. in the weak field limit. Therefore our results are limited to this low energy regime and will lose validity if we approach the Swinger limit $E_{S}$. After the ordinary wave breaks, we can not predict its behaviour in this approximation. 

As we have showed in \cite{KadlecovaKornBulanov2019}, the long--wavelength approximation breaks when the frequencies of the interacting waves, $\omega_{\gamma}$ and $\Omega$ become high enough as
\begin{equation}
\omega_{\gamma}\Omega > m^2_{e}c^4/\hbar^2,
\end{equation}
at this level the photon--photon interaction can result in creation of real electron--positron pairs via Breit--Wheeler process \cite{BreitWheeler}, in saturation of wave steepening and in the electromagnetic shock wave formation. Near the threshold, the electron--positron creation cross section has a form \cite{Electrodynamics,GouldSchreder},
\begin{equation}
\sigma_{\gamma\gamma\rightarrow ep}=\pi r_e^2 \sqrt{\frac{\hbar^2 \omega \Omega}{m_e^2 c^4}-1},
\label{eq:sigma-e-p}
\end{equation} 
where  $\omega_{\gamma}$ and $\Omega$ are the frequencies of high energy photons and low frequency counter--propagating electromagnetic waves, respectively, and $r_{e}=e^2/m_{e}c^2$ is the classical electron radius.

\begin{figure}[h]
\centering
\includegraphics[width=0.52\textwidth]{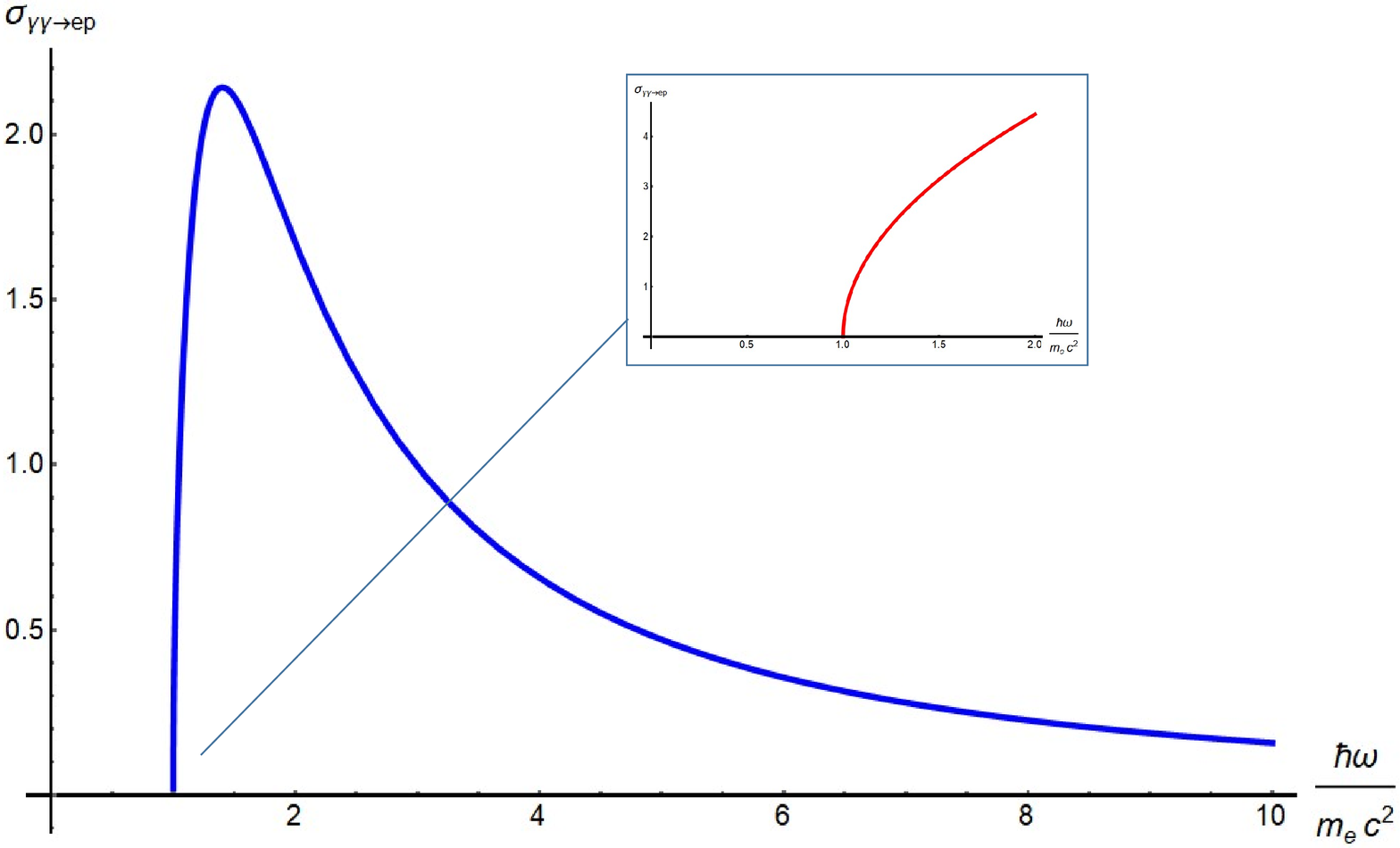}
\caption{\label{fig:CrossSectionEE} The cross section $\sigma_{\gamma\gamma\rightarrow ep}$ dependence on the photon energy $\hbar\omega/m_{e}c^2$. The detailed view on the graph near threshold is shown in the right insect. Reaching the energies for the electron--positron generation requires much less intense laser intensities than reaching Schwinger field $E_{S}$. }
\end{figure}

The cross section of the formation of an electron pair in the collision of two photons  is given by general formulae, \cite{Electrodynamics},
\begin{align}
\sigma_{\gamma\gamma\rightarrow ep}&=\frac{1}{2}\pi r_e^2(1-\beta_{e}^2)\times\nonumber\\
&\left\{(3-\beta_{e}^4)\log\left(\frac{1+\beta_{e}}{1-\beta_{e}}\right)-2\beta_{e}(2-\beta_{e}^2)\right\}{\rm d}\beta_{e},
\label{eq:sigma-e-e}
\end{align}
where 
\begin{equation}
\beta_{e}=\sqrt{1-\frac{m_{e}^2 c^4}{\hbar^2 \omega_{1}\omega_{2}}},
\end{equation}
the $\omega_{1}$ and $\omega_{2}$ are frequencies of the two colliding photons.
The cross section (\ref{eq:sigma-e-p}) is plotted  in Fig.~\ref{fig:CrossSectionEE} with detailed view on the area around the value one where the electron--poitron pair creation starts to appear.
When the shock wave has been formed we can find the electron pair number in a straightforward way using the energy--momentum conservation.
For a laser pulse of length $l_{pulse}$ and the beam profile $S'$, we can obtain the energy of the pulse as
\begin{equation}
\mathcal{E}_{pulse} \approx \frac{E^2_{pulse}S'l_{pulse}}{4\pi}=\frac{I_{pulse}S'l_{pulse}}{c},
\end{equation}
where $E_{pulse}$, $I_{pulse}$ are a typical field and intensity of the electromagnetic pulse,
then the number of the electron pairs and their creation rate are given by
\begin{equation}
N_{e^{\pm}}=\frac{\mathcal{E}_{pulse}}{2mc^2},\quad \frac{{\rm d} N_{e^{\pm}}}{{\rm d} t}=\frac{\mathcal{E}_{pulse}}{2mc^2}\frac{\delta v}{l_{pulse}},
\end{equation}
where we have denoted $\delta v=\bar{a}$. It can be shown that the creation rate of the electron pairs can be expressed as
\begin{equation}
N_{e^{\pm}}=\alpha\frac{I^3_{em}}{I_{S}^2},
\end{equation}
where $I_{em}$ is the intensity of electromagnetic field.

In our proposed model we target the lower energy region around the value one where the cross section curves start in Fig.~\ref{fig:CrossSectionEE}. 

The electron--positron pairs created at the electromagnetic shock wave front being accelerated by the 
electromagnetic wave emit gamma-ray photons which lead to the
electron--positron avalanche via the multi-photon Breit-Wheeler mechanism \cite{NikishovRitus} as discussed in Refs. \cite{BellKirk, Fedotov} (see also review article \cite{DiPizzaReview} and the literature cited therein).

Recently, in \cite{Yu2019}, a novel approach was developed to demonstrate the two-photon Breit--Wheeler process by using collimated and wide-bandwidth $\gamma$-ray pulses driven by $10$ PW lasers. The positron signal, which is roughly 100 times higher than the detection limit, can be measured by using the existing spectrometers. This approach, which could demonstrate the electron--positron pair creation process from two photons, would provide important tests for two-photon physics and other fundamental physical theories.


Additional terms corresponding to the dissipation effect due to viscosity and dispersion in the Eq.~(\ref{eq:finalOrdinary2}) can lead to saturation of high order harmonics \cite{Bulanov}. The dissipation effect can be described, for example, by additional term $\mu\partial_{xx}a^{(2)}=-\mu k^2 a^{(2)}$ on the left hand side of Eq.~(\ref{eq:secondorder}) as
\begin{align}
\partial_{t}a^{(2)}-\mu\partial_{xx}a^{(2)}=-\frac{bk}{2}\sin{[2k(x-vt)]},\label{eq:secondorderDiss}
\end{align}
assuming the dissipation effect have the same order as the second order in the wave amplitude perturbations.
 
The dispersion effect which is equivalent to the dependence of the phase
velocity on the wave number also can lead to saturation of the high harmonic
generation is described by additional term $\tau\partial_{xxx}a^{(2)}=-i\tau k^3 a^{(2)}$ on the left hand side of Eq.~(\ref{eq:secondorder}) as
\begin{align}
\partial_{t}a^{(2)}-\tau\partial_{xxx}a^{(2)}=-\frac{bk}{2}\sin{[2k(x-vt)]}.\label{eq:secondorderDiss1}
\end{align}
Both solutions of Eq.~(\ref{eq:secondorderDiss}, \ref{eq:secondorderDiss1}), amplitudes $a^{(2)}$, are less than the first harmonic amplitude in the limit $t\rightarrow\infty$. Saturation of the amplitude growth in the dispersive media appears due to the propagation velocity of the second harmonic being different from the background velocity $v$.


\section{\label{sec:conclusion} Conclusion}
In conclusion, we have presented an analytical method of solving the system of non--linear Heisenberg--Euler electrodynamics equations for a problem describing the
finite amplitude electromagnetic wave counter-propagating to the crossed electromagnetic field presented in \cite{KadlecovaKornBulanov2019}. We have used the weak field approximation to the sixth order in the field amplitude to include four and six photon interactions to study the singularity formation. It was shown that the non--linear field equations decouple for the ordinary wave case when we look for the solution in the form of a simple wave, i.e. Rieman wave, we have solved the equations exactly.
 
The solution has a form of non--linear wave equation for the relatively short wavelength pulse in the linear approximation and generalizes our previous result in \cite{KadlecovaKornBulanov2019}. The solution was analyzed by method of characteristics or by perturbation method and demonstrated in more detail that the solution describes high order harmonic generation, wave steepening and formation of a shock wave. The properties of the solution were discussed in detail, for example in the case of self--similar solutions with uniform (homogenious) deformation.

We analyze the electromagnetic wave steepening or wave breaking direction, it depends on the strength of the electromagnetic field $E_{0}$ (sign of $f'$) and has forward character for weak field and rarefaction shock wave character for stronger fields, as illustrated in Figs.~\ref{fig:shift} and \ref{fig:shift3D}.


In general, photon--photon scattering in a vacuum is governed by the dimensionless parameter $\alpha (I_{em}/I_S)$, as it concerns shock-like configuration formation, high order harmonics generation and the electron-positron and gamma ray flash at the electromagnetic shock wave front.

\begin{acknowledgments}
We thank Dr.~T. Pech\'{a}\v{c}ek for motivating discussions.
Supported by the project High Field Initiative (CZ$.02.1.01/0.0/0.0/15\_003/0000449$) from European Regional Development Fund.
\end{acknowledgments}

\bibliography{apssampELI1}

\end{document}